\documentstyle[prd,aps]{revtex}
\input epsf.tex
\def\DESepsf(#1 width #2){\epsfxsize=#2 \epsfbox{#1}}
\begin{document}
\draft
\title{Mixmaster Behavior in Inhomogeneous Cosmological Spacetimes}
\author{Marsha Weaver,$^{1,}$\thanks {Electronic address:
weaver@darkwing.uoregon.edu}
 James Isenberg,$^{2,}$\thanks {Electronic address: 
jim@newton.uoregon.edu}
 Beverly K. Berger$^{3,}$\thanks{Electronic address: berger@oakland.edu}}
\address{$^1$Department of Physics, University of Oregon,
Eugene, Oregon 97403 USA}
\address{$^2$Department of Mathematics, University of Oregon,
Eugene, Oregon 97403 USA}
\address{$^3$Department of Physics,
Oakland University, Rochester, MI 48309 USA}
\maketitle
\begin{abstract}
Numerical investigation of a class of inhomogeneous cosmological
spacetimes shows evidence that at a generic point in space the
evolution toward the initial singularity is asymptotically that of
a spatially homogeneous spacetime with Mixmaster behavior.  This
supports a long-standing conjecture due to Belinskii {\it et al}.
on the nature of the generic singularity in Einstein's equations.
\end{abstract}

If one assumes that our expanding universe can be described by
a spatially homogeneous and isotropic solution to Einstein's
equations, one can ``run the expansion backward'' to a hotter,
denser universe in the past.  Such an analysis leads to an
understanding of the cosmic microwave background and primordial
light element abundances.  Running a further finite time into the
past yields the big bang---a singularity characterized by infinite
density, temperature and gravitational tidal force.  While this
standard cosmological model accounts for observed features of
the universe, the reliability of its predictions about
features that have not been observed depends on the stability
of those predictions when the conditions of homogeneity and
isotropy are relaxed.

Einstein's equations allow for a rich variety of cosmological
spacetimes, by which we mean solutions that are
deterministic (contain a compact Cauchy
surface) and have a physically reasonable stress energy tensor
(one that satisfies the strong energy condition).  Powerful
theorems state that, generically, such spacetimes have an initial
singularity.  But the theorems do not describe the nature of the
singularity.  In the approach to the initial singularity in the
standard cosmological model, the kinetic energy of the isotropic
collapse (proportional to the square of the Hubble parameter)
dominates the spatial curvature.  A similar type of approach to
the singularity is found in the
Kasner spacetimes \cite{kasner}.  These vacuum
solutions are anisotropic, spatially homogeneous and spatially
flat (type I in the Bianchi classification of homogeneous spaces).
Since they are spatially flat, the spatial curvature terms are
absent from the evolution equations for these models, and
the kinetic energy of the anisotropic collapse drives the
approach to the singularity.  A cosmological spacetime is
said to have an asymptotically velocity term dominated (AVTD)
singularity if the evolution toward the singularity at each
spatial point approaches that of one of the Kasner spacetimes
or that of a nonvacuum Bianchi I spacetime with fixed Kasner
exponents\cite{im,els,explanation1}.  Another possible behavior near
the singularity is exemplified by the spatially homogeneous Mixmaster
spacetimes, in which the dynamics of the collapse to the singularity
is approximated by an infinite sequence of Kasner spacetimes with a
deterministically chaotic transition from one Kasner to the
next \cite{alt}.  In Bianchi type IX the
transitions are caused by ``bounces'' off a potential provided
by the spatial scalar curvature \cite{bkl,misner}.  In Bianchi
type VI$_0$ with magnetic field there
are bounces off a potential provided by the magnetic field in
addition to the curvature bounces \cite{lkw}.

While much progress has been made in understanding the homogeneous
case \cite{wh,ren2}, the behavior of spatially inhomogeneous solutions
to Einstein's equations near an initial singularity is largely unknown.
In a number of very limited classes of solutions (various classes of
spacetimes with 2-torus spatial symmetry and polarized vacuum spacetimes
with U(1) symmetry) there is strong evidence for AVTD
behavior \cite{im,inv,phen,ev}.  However, few expect AVTD behavior
to occur generally in cosmological spacetimes.  Rather the conjecture
has been that generically there is Mixmaster behavior, in which the
evolution toward the singularity at a generic spatial point approaches
that of one of the homogeneous Mixmaster
spacetimes \cite{bkl,quantum}. In a spatially inhomogeneous AVTD
spacetime the evolution at different spatial points will approach that of
different Kasner solutions.  In a spatially inhomogeneous spacetime that
has Mixmaster behavior the evolution at different spatial points will
approach that of different Mixmaster solutions.  Although these two
possibilities are mutually exclusive, in both cases the presence of the
inhomogeneity ceases to govern the dynamics asymptotically toward the
singularity.  This is a drastic assumption!  The space remains
inhomogeneous at all times, yet the effect of inhomogeneities on
the evolution becomes negligible.

Until now there has been no evidence that Mixmaster behavior
occurs in inhomogenous spacetimes.  We have obtained such evidence,
and discuss it in this letter.  We have focused on a class of
cosmological spacetimes that is an inhomogeneous generalization of
Bianchi type VI$_0$ with magnetic field.  Numerical study of
the evolution toward the singularity for a representative
sample of initial data shows that a regime consistent with
Mixmaster behavior is reached. It is impossible to follow
the evolution all the way to the singularity numerically. 
But analysis of the evolution equations under the conditions
that exist late in the numerical evolution show that the
regime will continue.

The spacetime manifold on which these solutions are
defined is $\Sigma^3 \times R $, where $\Sigma^3 $ is a solv-twisted  
2-torus bundle over the circle \cite{comment}.
While this manifold does not admit a
global two-torus action, it does admit the local group action
corresponding to Bianchi VI$_0$, which contains a local
two-torus action as a subgroup.  Spacetimes in the class we
studied have this local spatial two-torus symmetry and their
metrics can be written in the 3-torus Gowdy \cite{gow} form
with appropriate nonperiodic boundary conditions on some of
the metric coefficients. In particular, we can write the
metrics for this class   of spacetimes as
\begin{eqnarray}
g =& & - e^{{(\lambda (\theta,\tau) - 3 \tau) /
2}}\,d\tau^2 + e^{{(\lambda (\theta,\tau) +\mu
(\theta,\tau) +
\tau ) / 2}}\,d\theta^2 + \nonumber \\
& &e^{(P (\theta,\tau) - \tau)}  \Bigl[(dx + Q
(\theta,\tau)\,dy)^2 + e^{-2P (\theta,\tau)}\,dy^2 \Bigr].
\end{eqnarray}
The metric functions $\lambda$ and $\mu$ are periodic in
$\theta$.  $P (\theta + 2 \pi,\tau)=P (\theta,\tau)+ 2 \pi a$ and
$Q (\theta+ 2 \pi,\tau)=Q (\theta,\tau) e^{-2 \pi a}$.  Here $a$
is a constant determined by the manifold twist. 
When  $\lambda$ and $\mu$ are independent of $\theta$ and
$P (\theta,\tau)=P (\theta_0,\tau)+ a (\theta-\theta_0)$ and
$Q (\theta,\tau)=Q (\theta_0,\tau) e^{-a (\theta-\theta_0)}$ the
spacetime is locally homogeneous Bianchi VI$_0$.  As in
Bianchi VI$_0$ with magnetic field, we take the Maxwell
tensor to be
$F=B\,dx \wedge dy$. It then follows from the Einstein-Maxwell field  
equations that $B$ is necessarily constant
in space and time.  The nondynamical function $\mu$ is nonzero
only if the electromagnetic field is nonzero.  The time coordinate
$\tau$ has been defined (without loss of generality) so that
the singularity is at {\bf $\tau =  \infty$}.

The evolution equations for these solutions of the  
Einstein-Maxwell field equations can be derived from the
Hamiltonian density $H=H_1+H_2+H_3$,
\begin{eqnarray}
 H_1 &=& {1\over 4 \pi_\lambda}\Bigl[ \pi_P^2 +
e^{-2P} \pi_Q^2 \Bigr], \\
H_2 &=& {1\over 4 \pi_\lambda}\Bigl[e^{-2\tau} P'^2 +
e^{2(P-\tau)} Q'^2 \Bigr],\\
H_3 &=& 4 \pi_\lambda e^{{(\lambda+\tau)/ 2}} B^2.
\end{eqnarray} 
In addition, the following constraint equations
must be satisfied,
\begin{equation} 
\pi_P P' + \pi_Q Q' + \pi_\lambda \lambda' =
0, \quad \pi_\lambda = {1 \over 2} e^{\mu / 4},
\end{equation}
where $\pi_P$, $\pi_Q$ and $\pi_\lambda$ are the momenta conjugate to
$P$, $Q$ and $\lambda$.

This three hamiltonian form is useful for the following reason.
If any one of the three subhamiltonians is taken
by itself, and the other two ignored, the system is
exactly solvable.  Thus, after approximating the
continuous system by a discrete one, Suzuki's decomposition of
exponential operators \cite{suzuki} can be used to decompose
the evolution operator, with each piece exact.
The computer code we use for numerical evolution is based on this  
decomposition. It is an adaptation of a code used for the Gowdy
spacetimes \cite{inv}, and uses the fourth order decomposition and also fourth
order accurate representation of the spatial derivatives.

One can understand the structure of the numerically observed  
metric evolution for these spacetimes in terms of the evolving
relative dominance of the three subhamiltonians. Let us fix a
spatial point $p=(\theta_0,x_0,y_0)$.  Once the Mixmaster regime
is reached, then for most of the evolution  
towards the singularity ($\tau \to \infty$) at that point, we observe  
numerically  that $H_1$ dominates $H_2$ and $H_3$, and the metric  
evolution is essentially that of some Kasner spacetime.
Time intervals during which this happens are called Kasner  
epochs. For intermittent short periods,
either $H_2$ or $H_3$ (but never both at once)
also becomes significant.  These are the potentials that
cause the bounces.  When $H_1 + H_2$ is
dominant at $p$, the evolution is essentially as if $p$ were in
one of the vacuum Bianchi II (Taub \cite{taub}) spacetimes.
When $H_1 + H_3$ is dominant, the evolution is
essentially as if $p$ were in one of the
Bianchi I with magnetic field (Rosen \cite{rosen}) spacetimes.
Both the Taub and the Rosen solutions approach
one Kasner solution toward the singularity and another
Kasner solution in the opposite time direction
($\tau \to -\infty$).  Given a Kasner solution
there is no more than one Taub or Rosen
solution that approaches it as $\tau \to -\infty$.
So given a particular Kasner epoch at point $p$,
one knows which Taub or Rosen solution approximates the
next bounce toward the singularity.  This allows one to approximate
the sequence of Kasner epochs that will occur in a
given Mixmaster evolution.

To understand qualitatively why the bounces occur, let us assume  
that the functions $P$, $Q$, $\lambda$,
$\pi_P$, $\pi_Q$, $\pi_\lambda$ and their derivatives
develop in time in such a way that they do not  
counteract any explicit exponential decay in any of the terms in
$H$ or the resulting evolution
equations.  We shall call this ``assumption A.''
For example, if at some spacetime point, $(p, \tau)$,
one has the following (which we shall call ``the Kasner conditions''),
\begin{equation}
\tau \gg 0, \quad
\lambda + \tau \ll 0, \quad
P - \tau \ll 0,
\end{equation}
then assumption A implies that
at that point $H_1 \gg H_2$ and $H_1 \gg H_3$; and further it
implies that the terms in the evolution equations which are
derived from $H_1$ dominate those derived from $H_2$ and $H_3$.
The relative values of $H_1$, $H_2$ and $H_3$ accurately
monitor the relative importance of the terms derived from them
in the evolution towards the singularity because the exponential
factors control the growth and decay of the terms in which they
are present.  Without exception, our numerical results support
assumption A, and we assume it throughout the following.

Let us say that the Kasner conditions are satisfied at some point
$(p,\tau)$, so the evolution at $(p,\tau)$ is dominated by $H_1$.
If $H_2$ and $H_3$ were
zero, the evolution at $p$ would be exactly Kasner.
The quantity which determines which type of bounce  
will occur next is
$v = \sqrt{H_1/ \pi_\lambda}$.  One calculates  
(using assumption A along with the Kasner conditions) that
${dH_1 / d\tau} \approx 0$ and
${d\pi_\lambda / d\tau} \approx 0$, so during a given Kasner  
epoch at $p$, $v$ is essentially constant in time. We now  
argue that if $v < 1$ there will be a magnetic bounce in finite  
time, while if $v > 1 $ there will be a curvature  
bounce in finite time.

The reason $v$ is so important during a Kasner epoch in  
determining the next bounce is because $H_3 / H_1$ is controlled  
by $e^{\lambda + \tau}$ and $H_2 / H_1$ is controlled by  
$e^{P-\tau}$, and in turn $\lambda +\tau$ and $P-\tau$ are controlled  
by $v$. In particular, we find that during a Kasner epoch,
${d\lambda / d\tau} \approx - v^2$, so that 
${d (\lambda +\tau) / d\tau} \approx  ( 1- v^2)$.  Thus
if $v <1$, one has $e^{\lambda + \tau}$ increasing exponentially
with $\tau$ and therefore ${H_3 / H_1} \to {\cal O}(1)$
in finite time; while if $v >1$, it follows that $\lambda+ \tau$
decreases with $\tau$ and $H_3 / H_1$ stays small.
The evolution for $P$, governed by
${dP / d\tau}={\pi_P / 2 \pi_\lambda}$, is a bit more
complicated.  However, one finds $-v \leq {dP / d\tau} \leq +v$,
so if $v <1$, then ${d(P - \tau) / d\tau} < 0$ and
$H_2 / H_1$ must decrease.  Hence if $v <1$, a magnetic bounce
must occur.  Now consider the case $v >1$.  If $\pi_P > 0$
and therefore ${dP / d\tau} > 0$, then the Kasner evolution
leads to ${\pi_P / 2 \pi_\lambda} \to v$, and hence
${d(P - \tau) / d\tau} > 0$.  It follows that
${H_2 / H_1} \to {\cal O}(1)$ in finite time and there is
a curvature bounce.  If $\pi_P < 0$ and so ${dP / d\tau} < 0$,
then the Kasner evolution for $P$ has a single minimum after
which $\pi_P>0$ and the evolution proceeds to a curvature
bounce as just noted.  The evolution of $P$ past its minimum is
called a kinetic bounce.  The kinetic bounce, caused by
$e^{-2P} \pi_Q^2/ 4 \pi_\lambda$ in $H_1$, keeps $H_2$
from dying off in these spacetimes.  (See Figs. 1-3.)

What happens after a given bounce occurs? Since, as noted above,
a magnetic bounce is essentially a Rosen solution and a
curvature bounce is closely approximated by a Taub solution, one
can use the known features of those solutions to
determine the following \cite{lkw}.
1) After a magnetic bounce, induced by $v < 1$ at $p$, the metric  
evolution at $p$ returns to a Kasner epoch, this time with $v > 1$.
A curvature bounce will eventually follow. 2) After a curvature  
bounce induced by $ 1< v < 3$, one returns to a Kasner epoch with  
$v <1 $, so a magnetic bounce will follow. 3) After a curvature  
bounce induced by $v >3$, a Kasner epoch occurs with $v > 1$,
so another curvature bounce will follow.

The preceding characterizes the behavior at one point in
space.  Since $v$, $P$, and $\lambda$ are functions of $\theta$,
nearby points will in general not reach the end of a Kasner
epoch at exactly the same time.  (See Fig. 4.)

There are two different kinds of exceptions to the behavior just
described.  Neither violates assumption A, but the argument
that $H_2$ and $H_3$ continue to decay and then grow again
breaks down in the following ways.
First, there exist non-generic spacetimes in
the class we are considering which are not Mixmaster.  For instance,
if we set $B=0$, then $H_3$, which causes the magnetic
bounces, is missing and these spacetimes will be AVTD.

The second type of exception happens in a generic spacetime,
but only at isolated points, not on an open set in the
spacetime.  There are a number of field configurations
that prevent bounces at a point.  Some of these have
been seen in studies of the Gowdy spacetimes \cite{phen}:
$Q'=0$ when a curvature bounce would normally occur
prevents the curvature bounce, and $\pi_Q=0$ when a
kinetic bounce would normally occur prevents the
kinetic bounce (and hence the next curvature bounce if
$v>1$).  Others are new:  $v=1$ during a Kasner epoch
prevents both the curvature and magnetic type of bounce
and the Kasner epoch persists.
If $v=3$ during a Kasner epoch a curvature bounce will
occur, but the subsequent Kasner epoch has $v=1$.  If $\pi_P=0$
during a magnetic bounce the next curvature bounce is
prevented.  While more and more of these exceptional
points occur in a given spacetime as the evolution continues,
they will, for generic initial data, always be at isolated
values of $\theta$ \cite{explanation2}.

Our numerical study, combined with qualitative analysis
of the evolution equations, provides strong evidence that,
generically, spacetimes in this class exhibit Mixmaster
behavior.  While this class is spatially inhomogeneous, it
is still very restricted.  We predict that the particular
topology (which determines the boundary conditions on functions
of $\theta$) chosen is not necessary and that the generic
cosmological spacetime with local 2-torus symmetry and a magnetic
field perpendicular to the symmetry directions will also
have Mixmaster behavior.  But this is still a very restricted
class.  The question remains whether cosmological spacetimes
in general do indeed exhibit Mixmaster behavior, which would be
a surprising simplification of their evolution in the neighborhood
of the initial singularity, or whether some other possibility
in their evolution toward the initial singularity exists \cite{alt}.

This work was supported in part by a Federal Department of
Education Grant to the University of Oregon and by
NSF Grants PHY-9308177 to the University of Oregon and
PHY-9507313 to Oakland University.

\begin{figure}[bth]
\centerline{ \DESepsf(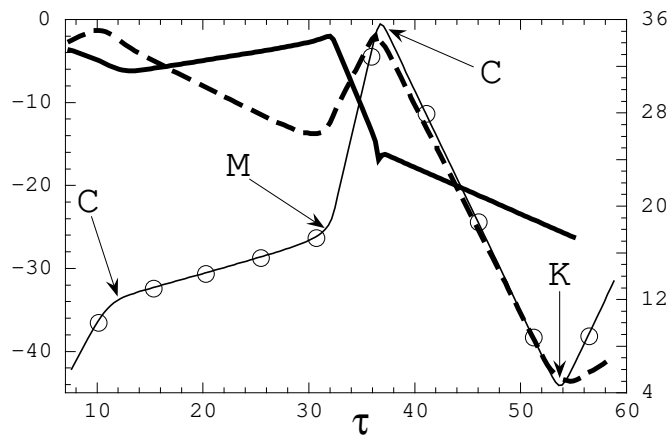 width 9 cm) }
\caption{Typical evolution at one spatial point.
The thick dashed line is $\log_{10}({H_2 \over H_1})$ and
the thick solid line is $\log_{10}({H_3 \over H_1})$ (left axis scale).
The thin solid line with circles is $P$ (right axis scale).
The bounces are labeled.  Note that $H_2 \approx H_1$
during the curvature bounces and $H_3 \approx H_1$ during
the magnetic bounce.  ${H_2 \over H_1}$ starts to grow
after the kinetic bounce.}
\label{1}
\end{figure}

\begin{figure}[bth]
\centerline{ \DESepsf(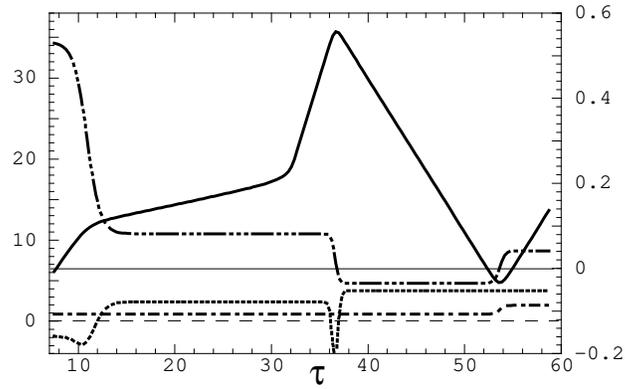 width 9 cm) }
\caption{Evolution at same spatial point as in Fig. 1.
The solid line is $P$
and the dotted line is $\pi_Q$
(left axis scale). The dash-single-dotted line is $Q$ and
the dash-triple-dotted line is $\pi_P$ (right axis scale).
Note that $P$ is essentially linear in $\tau$ except
during bounces.  The other functions are essentially
constant for most of the evolution.  This graph shows
to which type of bounce each function responds.}
\label{2}
\end{figure}

\begin{figure}[bth]
\centerline{ \DESepsf(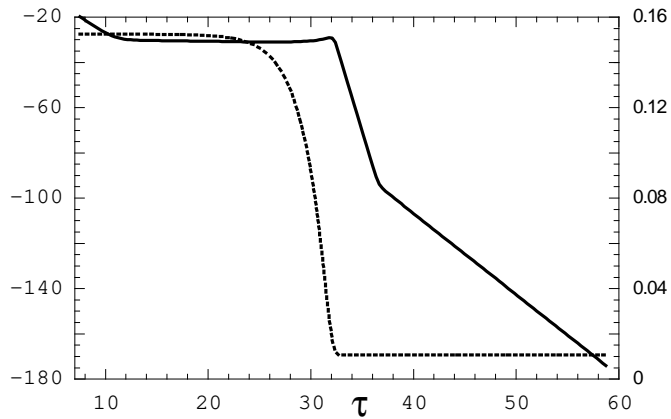 width 9 cm) }
\caption{Evolution at the same spatial point as in Figs. 1 and 2.
The solid line is $\lambda$ (left axis scale).
The dotted line is $\pi_\lambda$ (right axis scale).
Note that $\lambda$ is essentially linear in $\tau$ except
during bounces.}
\label{3}
\end{figure}

\begin{figure}[bth]
\centerline{ \DESepsf(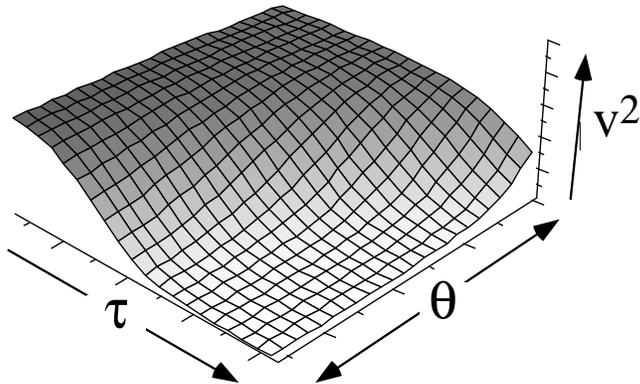 width 9 cm) }
\caption{This shows $v^2$ during the first curvature bounce of
Figs. 1-3 with neighboring $\theta$ values included.  $v^2$ is
essentially constant in time during a Kasner epoch and changes
during a curvature or a magnetic bounce.}
\label{4}   
\end{figure}


\begin{references}

\bibitem{kasner}
E. Kasner, Am. J. Math {\bf43}, 217 (1921).

\bibitem{im}
J. Isenberg and V. Moncrief, Ann. Phys. (N.Y.) {\bf199}, 84 (1990).

\bibitem{els}
D. Eardley, E. Liang, and R. Sachs, J. Math. Phys. {\bf13}, 99 (1972).

\bibitem{explanation1}  The Kasner exponents are the eigenvalues of
the extrinsic curvature, divided by the mean curvature.  In a
Kasner spacetime, the sum of the squares of the Kasner exponents
equals one.  In a nonvacuum Bianchi I spacetime the sum of the squares
of the Kasner exponents is less than one.
For examples of cosmological spacetimes which are AVTD but
whose evolution toward the singularity does not approach that of
a Kasner spacetime see Ref.\cite{wh} and A. D. Rendall,
J. Math. Phys. {\bf37}, 438 (1996).

\bibitem{alt} There may be other possiblilities.
See P. Breitenlohner, G. Lavrelashvili, and D. Maison, gr-qc/9711024;
E. E. Donets, D. V. Gal'tsov and M. Yu. Zotov, Phys. Rev. D {\bf56},
3459 (1997).

\bibitem{bkl} V. A. Belinskii, I. M. Khalatnikov, and E. M. Lifshitz,
Sov. Phys. Usp. {\bf 13},  745  (1971);  Adv. Phys. {\bf31}, 639 (1982).

\bibitem{misner} C. W. Misner, Phys. Rev. Lett. {\bf22}, 1071 (1969).

\bibitem{lkw} V. G. LeBlanc, D. Kerr, and J. Wainwright,
Classical Quantum Gravity {\bf12}, 513 (1995).

\bibitem{wh} J. Wainwright and L. Hsu,
Classical Quantum Gravity {\bf6}, 1409 (1989).

\bibitem{ren2} A. D. Rendall, Classical Quantum Gravity {\bf14},
2341 (1997); C. G. Hewitt and J. Wainwright, Classical
Quantum Gravity {\bf10}, 99 (1993).

\bibitem{inv}
B. K. Berger, V. Moncrief, Phys. Rev. D {\bf48}, 4676 (1993).

\bibitem{phen}
B. K. Berger and D. Garfinkle, Phys. Rev. D, 
to be published, gr-qc/9710102.

\bibitem{ev}  B. K. Berger and V. Moncrief, gr-qc/9801078;
S. Kichenassamy and A. D. Rendall, Classical Quantum Gravity,
to be published;
A. D. Rendall, Gen. Relativ. Gravit. {\bf27}, 213 (1995);
Classical Quantum Gravity {\bf 12}, 1517 (1995);
J. Isenberg and S. Kichenassamy, unpublished; 
P. T. Chru\'{s}ciel, {\it On Uniqueness in the Large of
Solutions of Einstein's Equations (``Strong Cosmic Censorship'')}
(Centre for Mathematics and its Applications, Australian National
University, Canberra, 1991).

\bibitem{quantum} Relating the ratio between the Hubble time and the
Planck time to the logarithm of the spatial volume in a typical Mixmaster
evolution shows that, independent of the definition of time, most
Mixmaster oscillations occur before the Planck time.  However, the spirit
of this field of inquiry is to determine the nature of the singularity
in classical general relativity.  One would expect that, just as in
electromagnetism, it will be important to understand the behavior
of classical solutions even if a quantum theory of gravity is found.

\bibitem{comment} Y. Fujiwara, H. Ishihara, and H. Kodama,
Classical Quantum Gravity {\bf10}, 859 (1993).

\bibitem{gow} R. Gowdy, Phys. Rev. Lett. {\bf 27},  826  (1971); Ann.
Phys. (N.Y.) {\bf83}, 203 (1974).

\bibitem{suzuki} M. Suzuki, Phys. Lett. A {\bf146}, 319 (1990).

\bibitem{taub}  A. H. Taub, Ann. Math. {\bf53}, 472 (1951).

\bibitem{rosen} G. Rosen, J. Math. Phys. {\bf3}, 313 (1962).

\bibitem{explanation2} The determination of exactly which
spatial point is the exceptional one is delicate.  At
any given time in the evolution, the function under
consideration will vanish (or equal 1) at one value of $\theta$.
The generic behavior is for the function to cross
0 (or 1) in a neighborhood of $\theta$.
The time derivative of that function is very small but nonzero,
so the point at which it vanishes will move in time.

\end{references}
\end{document}